\documentclass{kluwer}    % Specifies the document style.
\usepackage{graphicx,amssymb}

\newdisplay{guess}{Conjecture}

\newcommand{\HII}{H\,\textsc{ii}}
\newcommand{\HI}{H\,\textsc{i}}
\newcommand{\DI}{D\,\textsc{i}}
\newcommand{\FeI}{[Fe\,\textsc{i}]}
\newcommand{\FeII}{[Fe\,\textsc{ii}]}
\newcommand{\OI}{[O\,\textsc{i}]}
\newcommand{\OIII}{[O\,\textsc{iii}]}
\newcommand{\OIV}{[O\,\textsc{iv}]}
\newcommand{\NII}{[N\,\textsc{ii}]}
\newcommand{\NIII}{[N\,\textsc{iii}]}
\newcommand{\NeII}{[Ne\,\textsc{ii}]}
\newcommand{\NeIII}{[Ne\,\textsc{iii}]}
\newcommand{\NeV}{[Ne\,\textsc{v}]}

\newcommand{\SIII}{[S\,\textsc{iii}]}
\newcommand{\SIV}{[S\,\textsc{iv}]}
\newcommand{\ArII}{[Ar\,\textsc{ii}]}
\newcommand{\ArIII}{[Ar\,\textsc{iii}]}
\newcommand{\CII}{[C\,\textsc{ii}]}
\newcommand{\SiII}{[Si\,\textsc{ii}]}
\newcommand{\NiII}{[Ni\,\textsc{ii}]}
\newcommand{\micron}{$\mu$m}

\def\cmmt{~cm$^{-3}$}
\def\cmmd{~cm$^{-2}$}

\def\mum{$\mu$m}

\begin{document}
\begin{article}

\begin{opening}         
\title{High excitation ISM and gas} 
\author{E. \surname{Peeters}}  
\runningauthor{Peeters, Mart\'{\i}n-Hern\'{a}ndez, Rodr\'{\i}guez-Fern\'{a}ndez \& Tielens}
\runningtitle{High excitation ISM and gas}
\institute{NASA Ames Research Center, USA}
\author{N.L. \surname{Mart\'{\i}n-Hern\'{a}ndez}} 
\institute{Observatoire de Geneve, Switzerland}
\author{N.J. \surname{Rodr\'{\i}guez-Fern\'{a}ndez}} 
\institute{LUTH / LERMA - Observatoire de Paris, France}
\author{A.G.G.M. \surname{Tielens}} 
\institute{Kapteyn Astronomical Institute, The Netherlands}

\begin{abstract}
An overview is given of ISO results on regions of high excitation ISM
and gas, i.e. \HII\, regions, the Galactic Centre and Supernovae
Remnants. IR emission due to fine-structure lines, molecular hydrogen,
silicates, polycyclic aromatic hydrocarbons and dust are summarised,
their diagnostic capabilities illustrated and their implications
highlighted.

\end{abstract}

\end{opening}           

\section{Introduction}

Massive stars provide much of the radiative stellar energy of the
Milky Way.  Their copious amount of UV radiation has a great impact on
the surroundings of the massive star. Indeed, their UV radiation
dissociates molecules and dust in the interstellar medium (ISM) and
ionises hydrogen, creating large \HII\ regions.  In addition, their
powerful winds and supernova explosions provide most of the mechanical
energy of the Galaxy which dominate the structure of the
ISM. Furthermore, the nuclear reactions during their lifetime and
death scene synthesis most of the intermediate mass elements and
likely the r-process elements. Much of these nucleosynthetic products
may condense in the form of small dust grains. Indeed, supernova may
dominate the dust mass budget of the ISM. Because massive stars evolve
so fast, they are generally associated with the remnants of their
``cradle'' and are heavily enshrouded in dust and gas.  The gas and
dust around these stars absorbs most of the stellar luminosity which
is then re-emitted in the infrared. Hence, due to the high degree of
obscuration, massive stars can be best studied at wavelengths longer
than 2 \mum.

Offering unique combinations of wavelength coverage, sensitivity, and
spatial and spectral resolutions in the infrared spectral region, ISO
opened the infrared Universe. Its spectral coverage (from 2.3 -- 200
\mum) gives for the first time access to nearly all the atomic
fine-structure and hydrogen recombination lines in the infrared
range. In addition to the atomic lines, ISO revealed the shape and
strength of the dust continuum and several emission features.  This
presented an unprecedented view of the luminous but dusty and obscured
Universe, from the Galactic Centre to the most extincted regions of the
Milky Way and other galaxies.

This chapter reviews the major achievements of ISO on massive
stars. Sect.~\ref{hii} summarises the results on \HII\, regions, the
Galactic Centre is discussed in Sect.~\ref{gc}. Subsequently,
Sect.~\ref{sn} highlights the ISO results on supernova remnants.

\section{H\,{\footnotesize II} regions}
\label{hii}

The overall mid-IR (MIR) spectrum of \HII\ regions is dominated by a
dust continuum which rises strongly towards longer wavelengths
(Fig.\ref{spectrum}). On top of this continuum, there are a multitude
of fine-structure lines and hydrogen recombination lines. The
continuum emission is dominated by strong emission features at 3.3,
6.2, 7.7, 8.6 and 11.2 \mum, generally attributed to Polycyclic
Aromatic Hydrocarbons (PAHs). In some case, broad absorption features
due to simple molecules (H$_2$O, CO, CO$_2$) in an icy mantle and/or
narrow emission or absorption lines due to gaseous molecules (H$_2$,
H$_2$O, CO$_2$, OH, C$_2$H$_2$) are present.

\subsection{The line spectrum}

\subsubsection{The ionised gas}

\HII\ regions are prime targets to derive the present-day elemental
abundances of the ISM. They are bright and are characterised by a
large number of emission lines.  Optical studies of abundance
gradients in the Galaxy, however, present the problem that many \HII\
regions are highly obscured by dust lying close to the Galactic
plane. The arrival of the Kuiper Airborne Observatory
(e.g. \citeauthor{simpson95}, \citeyear{simpson95};
\citeauthor{afflerbach97}, \citeyear{afflerbach97};
\citeauthor{rudolph97}, \citeyear{rudolph97}) and the Infrared
Astronomical Satellite (e.g. \citeauthor{simpson90},
\citeyear{simpson90}) permitted measurements of elemental abundances
of embedded compact and ultra-compact \HII\ regions, obscured at
optical wavelengths but observable in the infrared, and gave access
for the first time to the central regions of the Galaxy.  Regarding
the determination of elemental abundances, the use of infrared
fine-structure lines present clear advantages with respect to the
optical lines (e.g. \citeauthor{rubin88}, \citeyear{rubin88};
\citeauthor{simpson95}, \citeyear{simpson95}): (1) they are attenuated
much less due to the presence of dust; (2) they are practically
insensitive to the precise temperature of the emitting gas since they
are emitted from levels with very low excitation energies; and (3) the
infrared range is the only wavelength regime to measure the dominant
form of nitrogen in highly excited \HII\ regions (N$^{++}$).

%%%%%%%%%%%%%%%%%%%%%%%%%%%%%%%%%%%%%%%%%%%%%%%%%%%%%%%%%%%%%%%%%%%%%%%%%
\begin{figure}[t!]
\center
\includegraphics[clip,height=6cm]{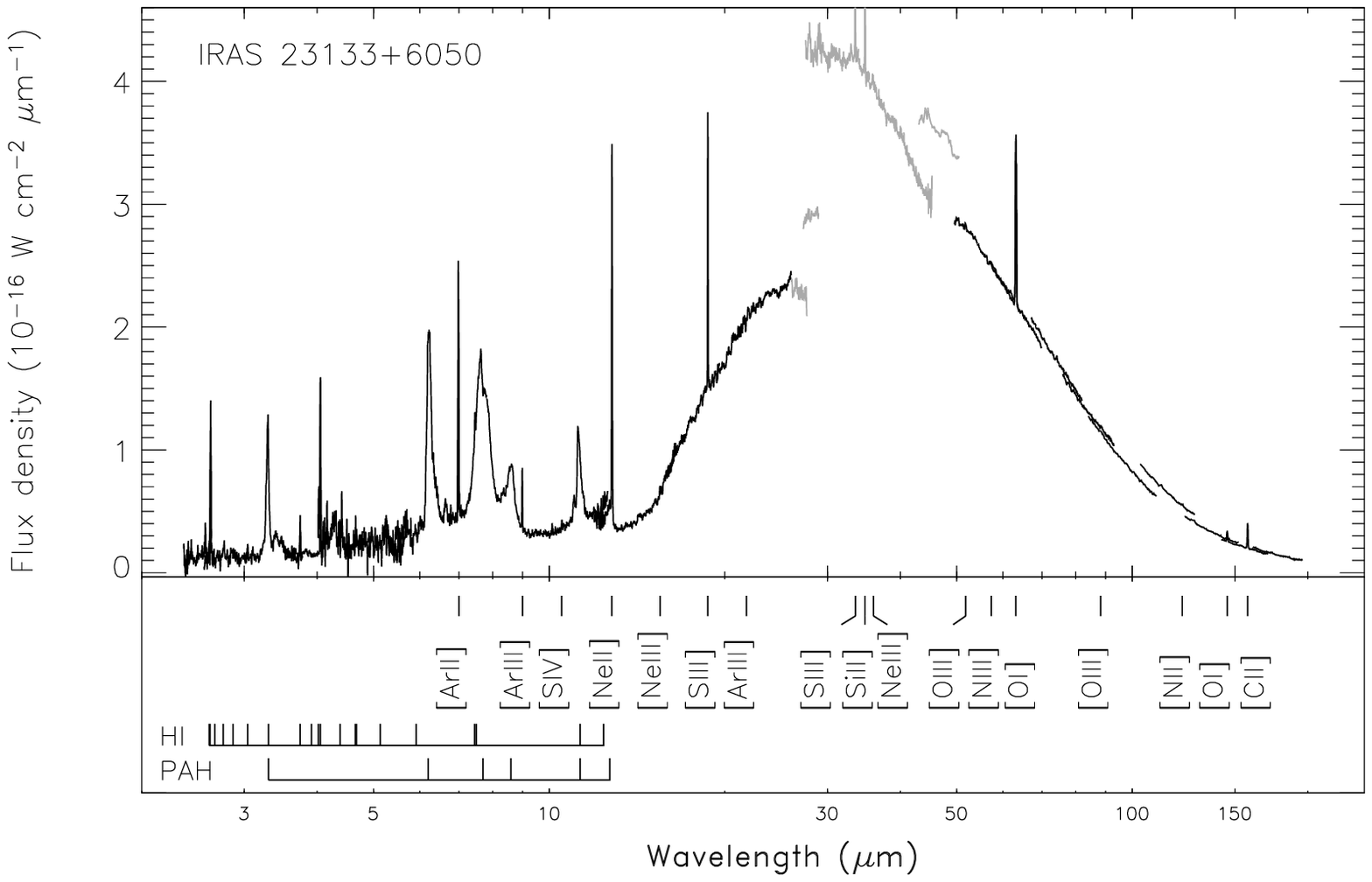}
\caption{The ISO-SWS/LWS spectra of a typical \HII\, region (Peeters et
al. 2002b).}
\label{spectrum}
\end{figure}
%%%%%%%%%%%%%%%%%%%%%%%%%%%%%%%%%%%%%%%%%%%%%%%%%%%%%%%%%%%%%%%%%%%%%%%%%

\begin{figure}[t!]
\center
\includegraphics[clip,width=5.5cm]{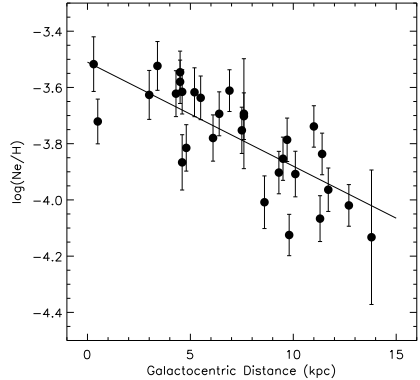}
\caption{Neon abundance as a function of the distance to the Galactic 
Centre (cf. Mart{\'{\i}}n-Hern{\' a}ndez et al., 2002a).}
\label{neon_abund}
\end{figure}

ISO provided a unique opportunity to measure the full spectral range
from 2.3 to 196~\micron\ of a large number of (ultra)compact \HII\
regions with relatively good spectral and spatial resolution
(e.g. \citeauthor{peeters02},
\citeyear{peeters02}). Fig.~\ref{spectrum} shows the combined SWS and
LWS spectrum of a typical \HII\ region. The spectrum is dominated by
recombination lines of H and fine-structure lines of C, N, O, Ne, S,
Ar and Si. The lines of \CII, \OI, and \SiII\ are produced by ions
with ionisation potentials lower than 13.6~eV and are thus expected to
be mostly emitted in the PDR surrounding the \HII\ region (see
Sect.~\ref{pdrlines}).  Two lines are present for some of the ions
(O$^{++}$, Ne$^{++}$ and S$^{++}$), providing a handle on the electron
density.  Typically, \OIII\ 52, 88~\micron\ densities towards Galactic
\HII\ regions range between $\sim$100 and 3000~cm$^{-3}$
(cf. \citeauthor{martin02}, \citeyear{martin02}). The use of the \SIII\
18.7, 33.5~\micron\ and \NeIII\ 15.5, 36.0~\micron\ line ratios
requires careful aperture corrections. When such corrections are
applied, the \SIII\ and \NeIII\ densities agree well with the \OIII\
densities within the errors (cf. \citeauthor{martin03},
\citeyear{martin03}).  N, Ne, Ar and S are observed in two different
ionisation stages, which enormously alleviates the problem of applying
ionisation correction factors (cf.  \citeauthor{martin02},
\citeyear{martin02}).  The ISO observations of \HII\ regions covered
the Galactic plane from the centre to a Galactocentric distance of
about 15~kpc, giving thus the possibility of investigating trends of
relative and absolute elemental abundances across a large part of the
Galactic disk. The gradients resulting from the full samples of
\citeauthor{martin02} \shortcite{martin02} and \citeauthor{giveon02}
\shortcite{giveon02} are $\Delta$log(Ne/H)$=-0.039\pm0.007$,
$\Delta$log(Ar/H)$=-0.047\pm0.007$, $\Delta$log(S/H)$=-0.023\pm0.014$
-- re-computed by \citeauthor{giveon02:note} \shortcite{giveon02:note}
applying the same extinction and $T_{\rm e}$ corrections to both data
sets -- and $\Delta$log(N/O)$=-0.056\pm0.009$~dex~kpc$^{-1}$.
Fig.~\ref{neon_abund} shows the neon abundance as a function of the
distance from the Galactic Centre.

%%%%%%%%%%%%%%%%%%%%%%%%%%%%%%%%%%%%%%%%%%%%%%%%%%%%%%%%%%%%%%%%%%%%%%%%%
%%%%%%%%%%%%%%%%%%%%%%%%%%%%%%%%%%%%%%%%%%%%%%%%%%%%%%%%%%%%%%%%%%%%%%%%%

%%%%%%%%%%%%%%%%%%%%%%%%%%%%%%%%%%%%%%%%%%%%%%%%%%%%%%%%%%%%%%%%%%%%%%%%%
\begin{figure}[t!]
\center
\includegraphics[clip,width=11.cm]{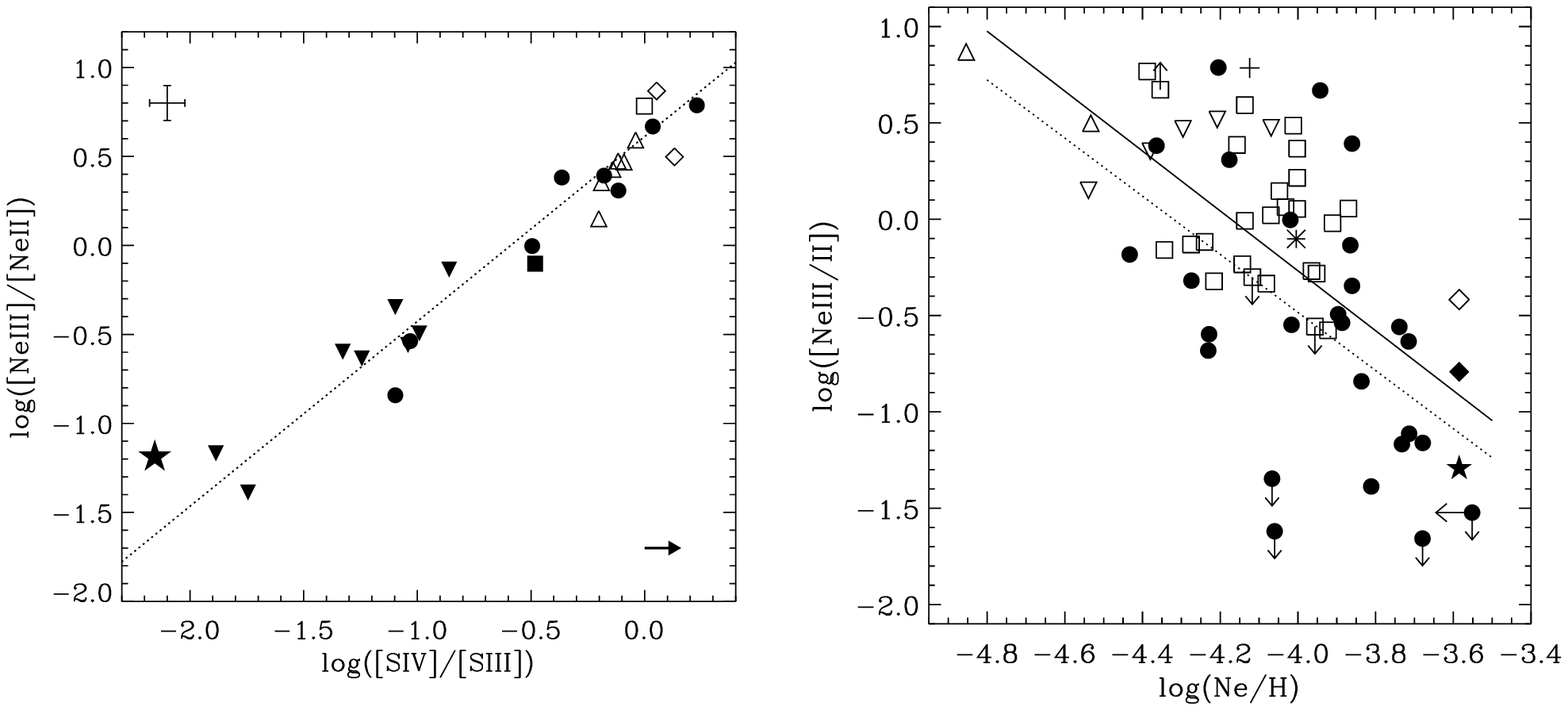}
\caption{Left: Relation between the \SIV/\SIII\ 10.5/18.7~\micron\ and
\NeIII/\NeII\ 15.5/12.8~\micron\ line ratios for a sample of 
\HII\ regions (cf. Mart{\'{\i}}n-Hern{\' a}ndez et al., 2002b). 
Indicated by various symbols are Galactic regions at
Galactocentric distance $R_{\rm gal} < 7$ kpc (solid triangles), 
Galactic regions at $R_{\rm gal} > 7$ kpc (solid circles), LMC regions
(open triangles, except 30\,Doradus, which is indicated by an open square)
and SMC regions (open diamonds). The positions of Sgr\,A$^\star$ 
(Lutz et al., 1996)  and the Orion nebula 
(Simpson et al., 1998) are indicated by a solid 
star and a solid square, respectively. The dotted line is a least squares
fit to the data. A typical error bar is given in the upper left corner.
The arrow in the lower right corner indicates the correction due
to an extinction $A_{\rm K}=2$ mag.
Right: Relation between the \NeIII/\NeII\ 15.5/12.8~\micron\ line ratio and the Ne/H 
elemental abundance for a combined sample of \HII\ regions
(Mart{\'{\i}}n-Hern{\' a}ndez et al., 2003). Indicated by various symbols
are Galactic \HII\ regions (solid circles), Sgr\,A$^\star$ (solid star),
the Pistol and the Sickle (open and solid diamonds, respectively; 
Rodr{\'{\i}}guez-Fern{\' a}ndez et al., 2001a), a sample of \HII\ regions
in M33 (open squares; Willner \& Nelson-Patel, 2002), 
LMC \HII\ regions (reverse open triangles, except 30\,Doradus, which is
plotted as a plus sign) and 2 regions in the SMC (open triangles).}
\label{metallicity1}
\end{figure}
%%%%%%%%%%%%%%%%%%%%%%%%%%%%%%%%%%%%%%%%%%%%%%%%%%%%%%%%%%%%%%%%%%%%%%%%%

The direct observation of two different ionisation stages not only
facilitates the determination of elemental abundances, but allows us
to probe the ionisation structure of the \HII\ regions and constrain
the stellar energy distribution (SED) of the ionising stars. Line
ratios such as \NIII/\NII\ 57/122~\micron, \NeIII/\NeII\
15.5/12.8~\micron, \SIV/\SIII\ 10.5/18.7~\micron\ and \ArIII/\ArII\
9.0/7.0~\micron\ probe the ionising stellar spectrum between 27.6 and
41~eV and depend on the shape of the SED and the nebular geometry (see
e.g. \citeauthor{morisset02}, \citeyear{morisset02}, who present a
detailed model of the well studied \HII\ region G29.96--0.02 based on
their infrared lines and \citeauthor{morisset04:2},
\citeyear{morisset04:2}, who compared predicted ionising spectra
against ISO observations of Galactic \HII\ regions). These line ratios
are found to correlate well with each other for the large sample of
\HII\ regions observed by ISO (cf. Fig.~\ref{metallicity1})
and to increase with Galactocentric distance
(cf. \citeauthor{giveon02}, \citeyear{giveon02}; \citeauthor{martin02}
\citeyear{martin02}). The observed \NeIII/\NeII\ and \SIV/\SIII\ line
ratios have been compared with diagnostic diagrams built from
extensive photoionisation model grids computed for single-star \HII\
regions using stellar atmosphere models from the WM-Basic code
\cite{pauldrach01} where the metallicities of both the star and the
nebula have been taking into account \cite{morisset04}. This
comparison finds no evidence of a gradient of the effective
temperature of the ionising stars with the Galactocentric distance,
attributing the observed increase of excitation with distance mainly
to the effect of the metallicity gradient on the SED (see also
\citeauthor{martin02:2}, \citeyear{martin02:2} and
\citeauthor{mokiem04}, \citeyear{mokiem04}). The relation between the
\NeIII/\NeII\ line ratio and the Ne/H elemental abundance for a
combined sample of Galactic and extra-galactic \HII\ regions is shown
in the left panel of Fig.~\ref{metallicity1}. As it is evident from this
figure, a clear correlation between excitation and metallicity exists,
albeit with a large scatter.

Extended emission of highly-ionised species (fine-structure lines of
N$^{+}$, N$^{++}$, O$^{++}$ and S$^{++}$) have been detected by LWS
and SWS (Mizutani~et~al.~2002, Okada~et al.~2003,
Goicoechea~et~al.~2004). This reveals the presence of extended, highly
ionised gas surrounding \HII\, regions and probably ionised by the
central O-star(s). The electron density has been derived from the {\sc
[O iii]} 52 to 88\,$\mu$m ratio, and values of a few 10 to a few 100
cm$^{-3}$ have been found. This extended, ionised gas, denser than the
galactic warm ionised medium would represent a new phase of the ISM as
stressed by Mizutani et al. (2002).

Extragalactic studies of \HII\ regions have been performed in the
Magellanic Clouds (cf. \citeauthor{vermeij02:1},
\citeyear{vermeij02:1}; \citeauthor{vermeij02:2},
\citeyear{vermeij02:2}) and M33 (\citeauthor{willner02},
\citeyear{willner02}). \HII\ regions in the Magellanic Clouds are
characterised by low metallicities and high excitation (see e.g.
Fig.~\ref{metallicity1}). 
Towards M33, the distribution of neon abundances as a function
of Galactocentric radius is best described as a step gradient, with a
slope of $-0.15$ dex from 0.7 to 4.0~kpc and $-0.35$ dex from 4.0 to
6.7~kpc.

\subsubsection{The photodissociation region}
\label{pdrlines}

Photodissociation Regions (PDRs) are regions where FUV ($6<h\nu<13.6$
eV) photons dissociate, ionise and heat neutral atomic/molecular gas
surrounding \HII\, regions (Tielens and Hollenbach 1985; Hollenbach
and Tielens 1999). The gas reaches temperatures in the range 100-1000
K, much warmer than the dust (10-100 K), and radiates its energy in
low energy atomic fine-structure lines, particularly the \OI\ 63
$\mu$m, \CII\ 157 $\mu$m, and \SiII\ 34 $\mu$m lines, and in pure
rotational molecular hydrogen lines (at wavelengths between 28 and 2
$\mu$m). In addition, the strong FUV flux pumps molecular hydrogen
molecules into excited vibrational states and their cascade produces
strong fluorescent ro-vibrational lines in the near-IR. The dust gives
rise to a continuum at long wavelengths ($\lambda>25\; \mu$m). In
addition, the IR spectrum of PDRs shows broad emission features at
3.3, 6.2, 7.7, 8.6, 11.2, and 12.7 $\mu$m due to fluorescent emission
of FUV pumped Polycyclic Aromatic Hydrocarbon (PAHs) molecules (see
Sect.~\ref{pahdust}). Here, we will focus on PDRs associated to \HII\
regions; other PDRs are extensively discussed elsewhere in this book.

Because of their luke warm temperatures, PDRs were only `discovered' when
the IR window was opened by the Kuiper Airborne Observatory in the
late 70ies and 80ies, but -- because of limited sensitivities -- these
studies focused on the brightest objects in the sky (the Orion bar;
Melnick et al., 1979; Storey et al., 1979, Russell et al., 1980, 1981).
The increased sensitivities of the SWS and LWS and, in particular, their
wide wavelength coverage allowed for the first time a systematic study of
the properties of PDRs. Much of this data are however still resting in the
archives. Except for the some of the ionic fine-structure lines and the \HI\
recombination lines, much of the IR emission characteristics of \HII\
regions (Fig.~\ref{spectrum}), originate in the associated PDRs.

%%%%%%%%%%%%%%%%%%%%%%%%%%%%%%%%%%%%%%%%%%%%%%%%%%%%%%%%%%%%%%%%%%%%%%%%%
\begin{figure}[t!]
\includegraphics[clip,width=5.5cm]{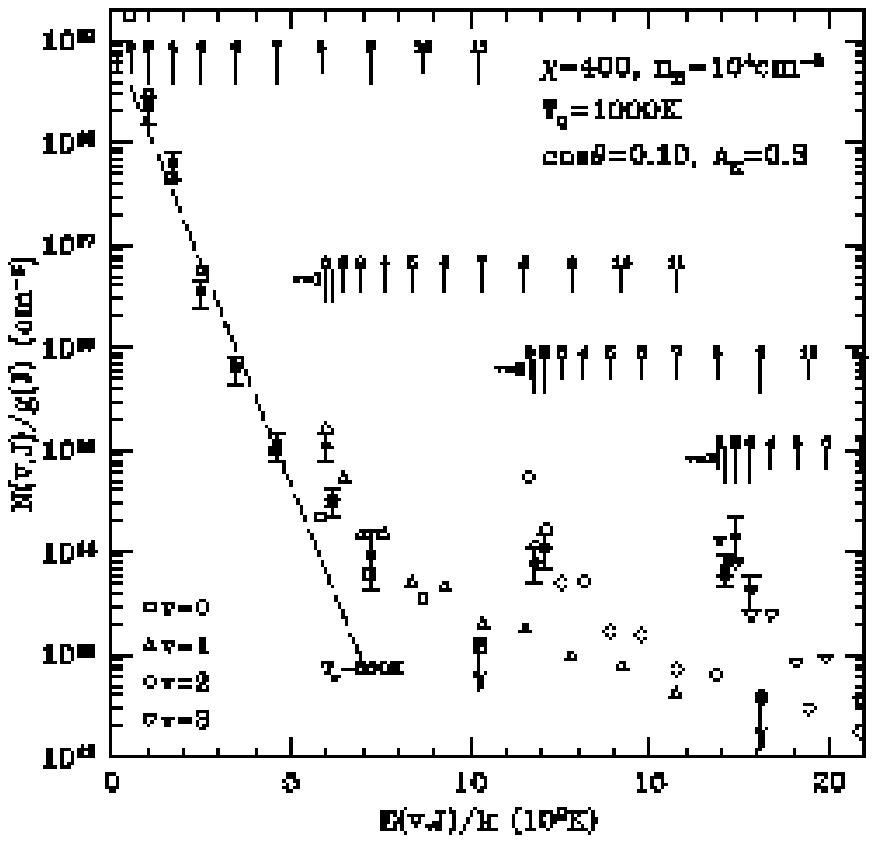}
\includegraphics[clip,width=5.5cm]{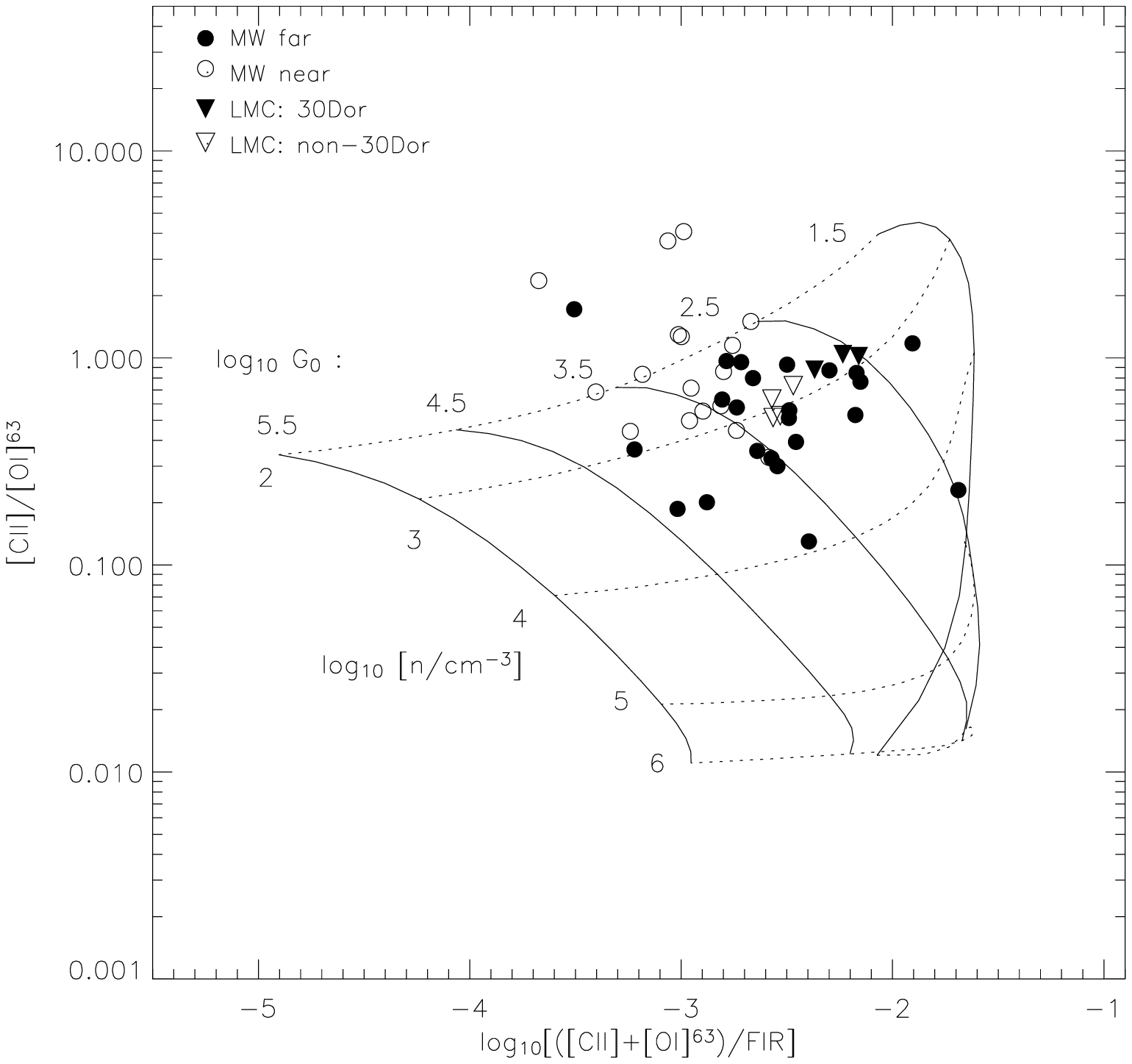}
\caption{{\bf Left :} The rotational diagram of molecular lines
observed towards the bright rim cloud, S140 (Timmermann et al.,
1996). Observed (filled symbols) and model calculated (open symbols)
column densities (ratioed to the statistical weight of the level
involved) are plotted as a function of the energy of the upper
level. A Boltzmann distribution will show a straight line on this
plot. The low lying pure rotational lines are consistent with gas at a
kinetic temperature of 500 K. The higher excitation temperature of the
vibrationally excited levels at high energies indicate the increased
importance of FUV pumping for the populations of these levels. The
models results are calculated for an incident FUV field of 400 times
the average interstellar radiation field and a gas density of $10^4$
cm$^{-3}$. A foreground extinction of 0.3 magnitudes in the K band has
been adopted. $T_o$ indicates the adopted peak temperature of the gas
in the PDR. {\bf Right :} A diagnostic diagram for PDRs based upon the
intensity ratio of the \CII\ 158 $\mu$m and the \OI\ 63 $\mu$m lines
and the overall cooling efficiency of the gas in the PDR. The lines
present the results of detailed model calculations for a range in
densities and incident FUV fields (Kaufman et al., 1999). The data are
taken from LWS observations of a sample of \HII\ regions (Peeters et
al., 2002; 2005; Vermeij  et al. 2002a).}
\label{h2}
\end{figure}
%%%%%%%%%%%%%%%%%%%%%%%%%%%%%%%%%%%%%%%%%%%%%%%%%%%%%%%%%%%%%%%%%%%%%%%%%

The SWS is particularly particularly suited for observations of the
pure rotational lines of molecular hydrogen. These lines provide a
direct handle on the physical properties of the emitting gas. Because
of the small Einstein A's of the associated quadrupole transitions,
the lowest levels of H$_2$ have very low critical densities and hence
their populations are in local thermodynamic equilibrium for typical
PDR conditions (Burton et al., 1992). A rotational diagram provides
then directly the temperature and column density of the emitting
gas. This is illustrated in Fig.~\ref{h2}, right panel, for the PDR
associated with the blister \HII\ region, S140 \cite{Timmermann}. The
optically bright rim is produced by the ionisation of a dense clump in
the molecular cloud, L1202/1204 by the nearby B0.5 star, HD 211880
(Blair et al., 1978; Evans et al., 1987). The SWS spectrum of this
source shows pure rotational H$_2$ lines up to $0-0\; S(9)$. The
column density and temperature derived from these observations are
$N\simeq 10^{20}$ cm$^{-2}$ and $T_{rot}=500$ K
\cite{Timmermann}. Similar studies of the pure rotational lines in the
spectra of S106 and Cep A resulted in excitation temperatures of 500
and 700 K \cite{vandenAncker, Wright}.

In addition, SWS has observed several ro-vibrational lines in S140
belonging to the $v=$1--0, 2--1, and 3--2 series. These levels are
populated by FUV pumping.  Fig.~\ref{h2}, right panel, compares the
observations with a detailed model for the collisional and FUV-pumping
excitation of H$_2$ in this source \cite{Timmermann}. Good agreement
is obtained for a density of $10^4$ cm$^{-3}$ in the PDR and an
incident FUV flux of $\simeq$400~Habing. These values are consistent
with estimates based upon pressure equilibrium across the ionisation
front coupled with the ionised gas density in the associated bright
rim and the properties of the illuminating star.

While this agreement between observations and models of H$_2$
excitation is very comforting (Fig.~\ref{h2}, right panel), PDR models
have a problem in matching the observed kinetic temperatures and
column densities derived from the low lying pure rotational lines. In
particular, for such low FUV fields and densities as derived for S140,
the calculated gas temperatures in the PDR are much lower than the
observed 500 K (e.g. Draine \& Bertoldi, 1999). This problem seems to
be more general, although a detailed comparison of observations and
models for regions spanning a wide range in physical properties is
still wanting.  Nevertheless, it seems that PDR models may be missing
an essential part of the energetic coupling between the gas and the
illuminating stellar FUV radiation field.

The far-IR atomic fine-structure lines provide important probes of the
physical conditions in PDRs. However, at present, little has been done
to harvest the data provided by the LWS. The \OI\ 63 $\mu$m and \CII\
157 $\mu$m lines are the dominant coolants of PDRs (Tielens and
Hollenbach 1985). Hence, the summed flux of these lines measures the
total heating of PDR gas by the FUV radiation field.  Comparison with
the total incident FUV flux, as measured by the far-IR dust continuum
provides a measure of the heating efficiency of PDRs. The levels
involved in these two transitions have very different critical
densities ($\sim 10^5$ and $3\times 10^3$ cm$^{-3}$) and excitation
energies (92 and 228 K). Not coincidentally, these values are
precisely the range expected for gas in PDRs and, hence, the ratio of
these lines has evolved to a major diagnostic of the properties of
PDRs. This is illustrated in Fig.~\ref{h2}, right panel, where the
results of PDR models for this line ratio are plotted against the
calculated ratio of the cooling lines as a function of density and
incident FUV field (Kaufman et al., 1999). Over much of the relevant
range of these observables, the model result segregate out well and
hence this diagram is very useful for the analysis of the conditions
in PDRs.

Observations towards a sample of \HII\ regions are also shown in
Fig.~\ref{h2}, left panel.  The conditions in these regions span a
range in density, 30-$3\times10^3$ cm$^{-3}$ and incident FUV fields,
$3\times10^2-3\times10^4$ Habing.  These values are within the range
expected for PDRs associated with (evolved) compact \HII\ regions. In
addition, there seems to be a trend in the distribution of the
observed points with location in the galaxy or, equivalently,
metallicity. That is regions with lower metallicity seem to be
characterised with a higher heating efficiency and lower \CII/\OI\
ratio than regions with higher metallicity. If the \OI\ line is the
dominant cooling line, a decrease in the metallicity is expected to
result in a decreased \CII/\OI ratio (Tielens and Hollenbach
1985). However, the decrease in the heating rate with increasing
metallicity is not expected. Further analysis of these types of
observations is clearly warranted.

The fine-structure lines observed towards several individual high mass
star forming regions have been analysed in detail. Analysis of the
observations of W49N show that the PDR is illuminated by an intense
FUV field (G$_o$=3$\times$10$^5$). The density and temperature are,
however, quite moderate ($n=10^4$\,cm$^{-3}$, $T=130$K). Reflecting
this high FUV field and low density, the observed heating efficiency
is comparatively low, $\sim10^{-4}$ \cite{Vastel:01}. Towards S106IR,
a hot (200-500K) dense (n\,$>$\,3x10$^5$\,cm$^{-3}$) gas component is
present. Cooler gas is associated with the bulk of the emission of the
molecular cloud and is characterised by 2 emission peaks which have
densities of 10$^5$ cm$^{-3}$ and are illuminated by a radiation
field, G$_0$, ranging from 10$^2$ to 10$^{3.5}$
\cite{Schneider:pdr}. The \HII\ region S~125 is modelled
self-consistently with a 2-dimensional geometrical blister model. In
order to fit the spatial profile, a systematic increase of the gas
temperature along the PDR boundary with decreasing distance from the
ionising star is necessary, which is not readily understood within
present-day PDR models \cite{Aannestad:line}.  Despite the rich
concentration of massive stars in the stellar cluster, Trumpler 14, in
the Carina nebula, the physical conditions in the associated PDR are
much less extreme than for Orion or M17 \cite{Brooks}. The data for
the PDR in NGC 2024, on the other hand, are consistent with emission
from dense ($n\simeq 10^6$ cm$^{-3}$), coolish ($T\simeq 100$ K)
clumps \cite{Giannini}.

Finally, the lowest pure rotational $J=1-0$ transition of the HD molecule
at 112 $\mu$m was detected towards the Orion Bar using the LWS on ISO
\cite{Wright:99}. The observations imply a total HD column density of
$\left(3\pm 0.8\right)\times 10^{17}$ cm$^{-2}$ for adopted temperatures
in the range 85--300 K. This corresponds to an elemental D/H abundance
ratio of $\left(2\pm 0.6\right) \times 10^{-5}$, comparable to that
derived from \DI\ and \HI\ ultraviolet absorption measurements in the solar
neighbourhood.

\subsection{PAHs and dust}
\label{pahdust}

Besides the well-known UIR bands at 3.3, 6.2, 7.6/7.8, 8.6, 11.2 and
12.7 \mum, and very broad structures present underneath these bands,
many discrete weaker bands and subcomponents can be found at 3.4, 3.5,
5.2, 5.7, 6.0, 6.6, 7.2--7.4, 8.2, 10.8, 11.0, 12.0, 13.2 and 14.5
\mum. ISO extended the observable wavelength range and found there new
feature at 16.4~$\mu$m and 17.4 \mum\, as well as a weak, variable,
emission plateau between 15 and 20~$\mu$m \cite{Moutou:16,
  VanKerckhoven, VanKerckhoven:phd}. It should be emphasized that not
all sources show all these emission features at the same time. 

\begin{figure}[t!]
\center
\includegraphics[clip, width=6cm, angle=90]{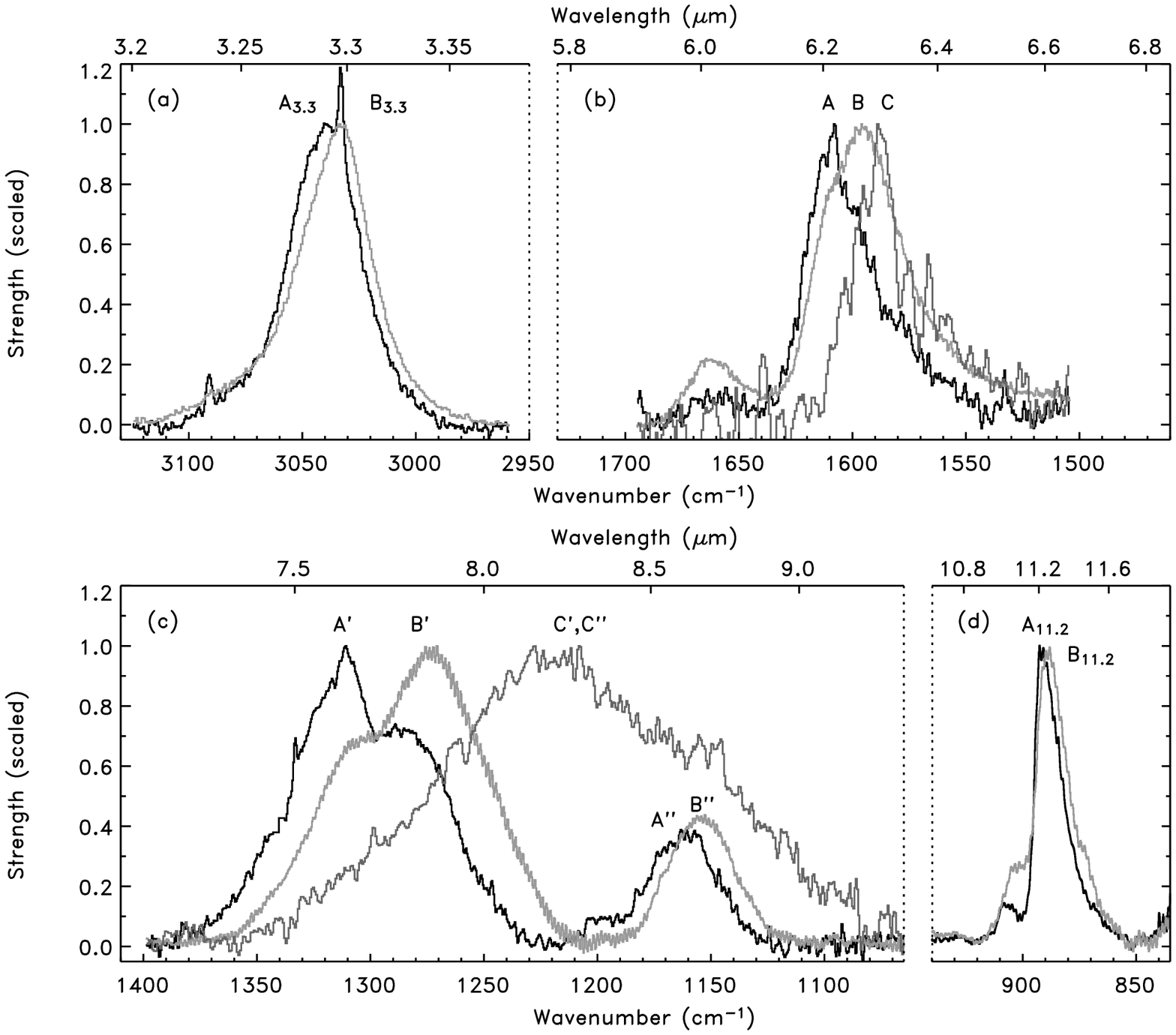}
\includegraphics[clip, width=4.cm,height=4.cm]{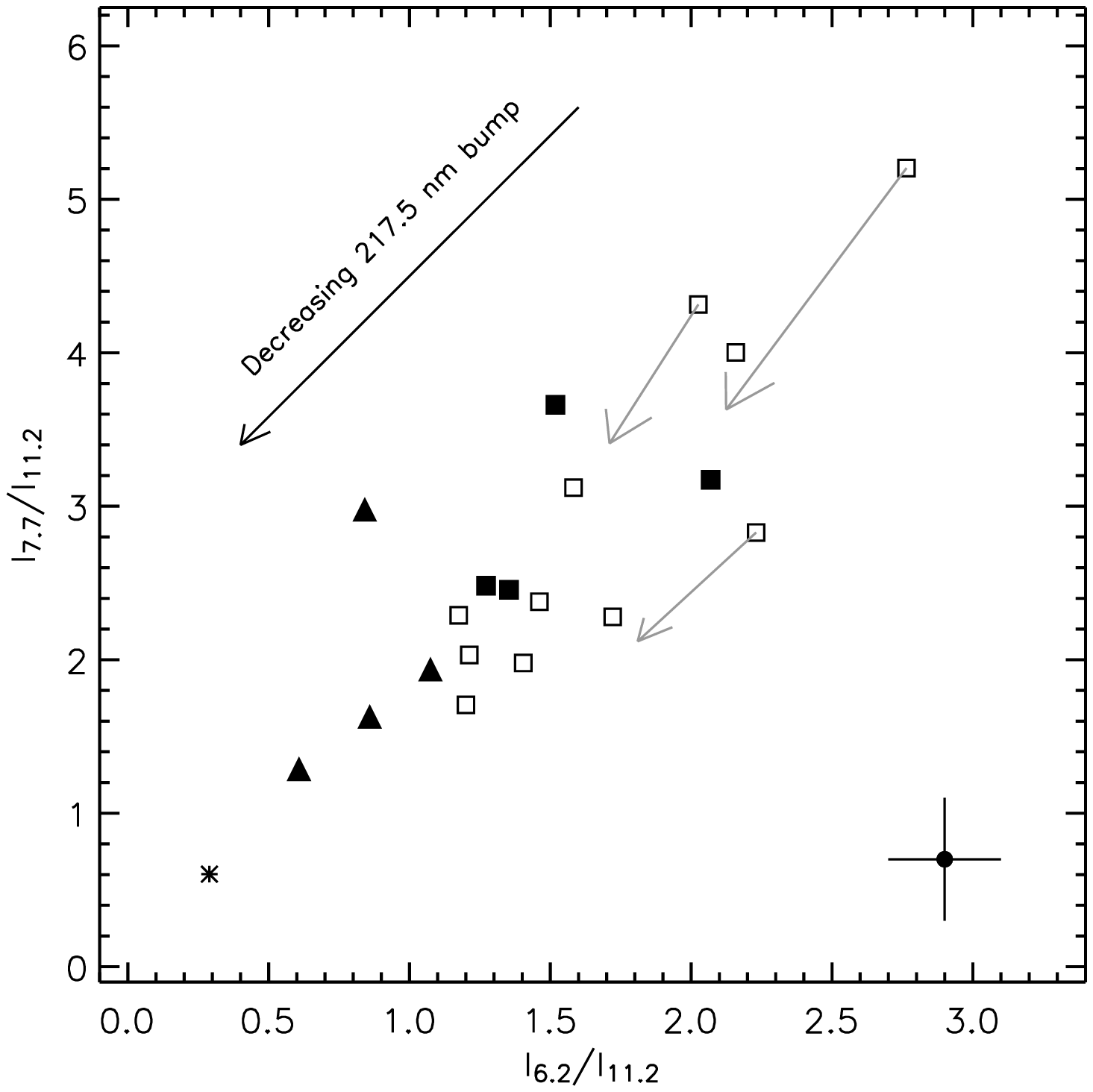}
\caption{Left: Source to source variations for the 3--12~$\mu$m UIR
  bands. Profile $A$ represent \HII\ regions, Reflection Nebulae
  and the ISM, profile $B$ isolated HAeBe stars and most PNe and
  profile C 2 post-AGB stars (Peeters et al. 2002a; van Diedenhoven et
  al. 2004). Right : PAH band strength ratios for \HII\
  regions. Galactic sources are shown as $\Box$, non-30~Dor sources as
  $\blacksquare$, 30 Dor pointings as $\blacktriangle$ and the SMC
  B1$\sharp$1 molecular cloud (Reach et al. 2000a)  by an asterisk (Vermeij
  et al. 2002).}
\label{pahs}
\end{figure}

ISO also allowed, for the first time, a systematic analysis of the UIR
bands in a wide variety of environments.  It is now firmly established
that the detailed characteristics (intensity, peak position, profile)
of the UIR features vary from source to source and also spatially
within extended sources (for a recent review, see
\citeauthor{Peeters:colorado}, \citeyear{Peeters:colorado}).  ISO-SWS
spectra indicated that the UIR band profiles and positions of all
\HII\ regions are equal to each other and to those observed in
Reflection Nebulae (RNe), non-isolated Herbig AeBe stars and the ISM,
while they are clearly distinct from those found around evolved stars
and isolated Herbig AeBe stars (Fig.~\ref{pahs}, Peeters et al. 2002a,
van Diedenhoven et al. 2004). However, spatial variations within RNe
\cite{Bregman} and evolved stars (\citeauthor{Kerr}, \citeyear{Kerr};
\citeauthor{Song}, \citeyear{Song}; \citeauthor{Miyata},
\citeyear{Miyata}) are observed suggesting that in all sources profile
variations occur on a small spatial scale. In addition, - in contrast
to what the integrated spectra of sources might suggest - the profiles
are not unique to certain object types.  In contrast, their relative
strength in \HII\ regions varies both from source to source and within
sources (Fig.~\ref{pahs}).  Integrated spectra of sources show
variations in the relative strength of the CC modes (6--9~$\mu$m
range) relative to the CH modes (at 3.3 and 11.2~$\mu$m), as is the
11.2/12.7 band ratio while the 3.3~$\mu$m feature correlates with the
11.2~$\mu$m feature and the 6.2~$\mu$m feature with the 7.7~$\mu$m
feature (Fig~\ref{pahs}, \citeauthor{Roelfsema}, \citeyear{Roelfsema};
\citeauthor{Verstraete:m17}, \citeyear{Verstraete:m17};
\citeauthor{Hony}, \citeyear{Hony}; \citeauthor{Vermeij:pah},
\citeyear{Vermeij:pah}).  These variations are directly related to
object type \cite{Hony} and to the local environment (Fig~\ref{pahs},
\citeauthor{Verstraete:m17}, \citeyear{Verstraete:m17};
\citeauthor{Vermeij:pah}, \citeyear{Vermeij:pah}).  Within extended
sources, additional variations are found, nicely demonstrated by that
of the 6.2/7.7 band ratio in S106 \cite{Joblin}.  Within a single
object, variations in G$_0$ are important in driving the UIR emission
spectrum \cite{Onaka, Bregman}. However, the source to source
variation does not seem to follow G$_0$ \cite{Verstraete:prof, Hony,
Peeters:prof, Vermeij:pah, vanDiedenhoven, Peeters:gal}.

%%%%%%%%%%%%%%%%%%%%%%%%%%%%%%%%%%%%%%%%%%%%%%%%%%%%%%%%%%%%%%%%%%%%%%%%%
%%%%%%%%%%%%%%%%%%%%%%%%%%%%%%%%%%%%%%%%%%%%%%%%%%%%%%%%%%%%%%%%%%%%%%%%%

The observed variations of the UIR bands provide direct clues to the
physical and chemical properties of their carriers, the local physical
conditions and/or the local history of the
emitting population.
These spectral variations have been attributed to a range of different
physical or chemical characteristics of the carriers, including charge
state, anharmonicity, different subcomponents with variable strength,
family of related species with varying composition,
substituted/complexed PAHs, isotope variations, clustering, molecular
structure, molecular composition etc. \cite{Verstraete:m17,
Joblin, Verstraete:prof, Hony, Pech, Peeters:prof, Wada, Song,
Bregman, Hudgins, vanDiedenhoven}. In particular, the 
variations in the 11.2/6.2 and 3.3/6.2 \mum\, ratios likely reflect
variations in the average ionization state of the emitting PAHs. In
contrast, the variations in the peak position of the 6.2 \mum\, band
point towards variations in the molecular structure of the emitting
species perhaps related to the incorporation of N into the ring
structure of the PAHs.

Since the UIR bands are omnipresent, they serve as important probes
of the different emission zones.  For example, the presence and
strength of the UIR bands are generally thought to trace star
formation and so they are used as qualitative and quantitative
diagnostics of the physical processes powering Galactic nuclei (see
this issue). In addition, deuterated PAHs are tentatively detected at
4.4 and 4.65 \mum\, at high abundances in the Orion Bar and M17
\cite{Peeters:pad}. If born out, they can serve as probes of PAH D/H
ratio in various regions. 

Several studies were devoted to the spatial distribution of the
different emission components \cite{Cesarsky:m17, Verstraete:m17,
Pilbratt, Crete, Klein, Zavagno, Urquhart, Verma}. It is
overwhelmingly clear that the UIR bands peak in the PDR, though closer
to the ionized star with respect to H$_2$, while the
strong dust continuum peaks towards the \HII\ region. Inside the \HII\
region, where UIR bands are weak or absent, remains a broad underlying
emission band which is attributed to very small grains (VSG,
\citeauthor{Cesarsky:m17}, \citeyear{Cesarsky:m17}; \citeauthor{Jones}
\citeyear{Jones}). The spatial distribution of the UIR bands is also
displaced from that of the ERE emission in Sh~152 \cite{Darbon},
similar to the Red Rectangle \cite{Kerr}. It is clear that the carrier
of the UIR bands is not also responsible for the ERE.  Moreover,
because the ERE emission can sometimes be traced to inside the ionized
gas and recalling that the PAHs do not survive in the \HII\, region,
the carrier of the ERE is unlikely to be molecular in origin.

The dust/UIR band emission has been modeled for several sources.
\citeauthor{Crete} (\citeyear{Crete}) modeled M17 using a
2-component model composed of PAHs as the carrier of the UIR
bands and VSG responsible for the ``warm'' continuum and a 2-component
model composed of coal dust emission and big grain emission. In the
frame of the coal dust model, the observations required nano-sized,
transiently heated particles. The IR emission in the M17-SW \HII\
region and Orion is modeled with a mixture of amorphous carbons,
silicates and possibly crystalline silicates (see below,
\citeauthor{Jones}, \citeyear{Jones}; \citeauthor{Cesarsky:m17}
\citeyear{Cesarsky:m17}). PAHs seem to be depleted inside the
\HII\, region consistent with their destruction in the intense
radiation field of M17 and Orion \cite{Jones, Cesarsky:silem,
Crete}. Modelling of the FIR emission of Sh~125 also indicates the
dust to be severely depleted in the \HII\ region while normal dust/gas
ratio is observed in the PDR, but with a major fraction in the form of
VSG \cite{Aannestad}.

ISO also increased the number of \HII\ regions showing silicate in
emission \cite{Cesarsky:silem, peeters02, Peeters:phd} and detected
several new dust features :
 
1) A very strong, broad 8.6 \mum\, band is detected towards
the M17-SW \HII\ region and 3 compact \HII\, regions, clearly distinct
from the ``classical'' 8.6 \mum\, PAH band and originating
inside the \HII\ regions \cite{Roelfsema, Cesarsky:m17,
Verstraete:m17, Peeters:prof, Peeters:89}. Detailed analysis reveals
the feature having a profile peaking at 8.9 \mum\, with a FWHM $\sim$1
\mum\, and likely not being related to PAH emission
\cite{Peeters:89}.

2) A broad 22 \mum\, feature is detected in the M17-SW \HII\ region
\cite{Jones}, the Carina Nebula and two starburst galaxies
\cite{Chan}. The feature shape is similar to that of the 22 \mum\,
band observed in Cassiopeia A indicating that supernovae are probably
the dominant production sources \cite{Chan}.

3) Several (possible) detections of crystalline silicates associated
with \HII\ regions are reported, i.e. (i) emission at 34 \mum\, in the
M17 \HII\ region attributed to Mg-rich crystalline olivine
\cite{Jones}; (ii) emission bands at 9.6 \mum\, in $\theta^2$ Ori A
and at $\sim$ 15-20 \mum, $\sim$ 20-28 \mum\, and longward of 32
\mum\, in the Orion \HII\ region attributed to crystalline Mg-rich
silicate fosterite \cite{Cesarsky:silem}. The latter emission bands
are however a combination of instrumental effects and the PAH 15-20 \mum\,
plateau (Kemper et al., in preparation); (iii) a band at 65 \mum\, in
the Carina Nebula and Sh~171 attributed to diopside (Ca-bearing
crystalline silicate - Onaka \& Okada, 2003). An upper-limit for the
crystallinity of the silicates in the ISM is derived being a few \%
\cite{Jones}, less than 5\% \cite{Li} and less than 0.4\%
\cite{Kemper04}.

4) To end, a broad emission feature is found at 100 \mum\, in the
Carina Nebula and Sh~171, possibly due to carbon onion grains
\cite{Onaka:03}. A similar broad feature around 90 \mum\, is reported
in evolved stars and a low mass protostar, attributed to
calcite, a Ca-bearing carbonate mineral \cite{Ceccarelli, Kemper}.

\section{The Galactic Centre}
\label{gc}

The 500 central pc of the Galaxy (hereafter Galactic center, GC) are
an extended source of emission at all wavelengths from radio to
$\gamma$-rays (see \citeauthor{Mezger:96}, \citeyear{Mezger:96} for a
review). The non-thermal radio emission as well as $\gamma$ and X-rays
observations show the presence of a very hot plasma ($10^7-10^8$ K)
and a recent episode of nucleosynthesis ($\sim10^6$ yr ago) that
suggest an event of violent star formation in the recent past. On the
other hand, the cold medium in the GC is mainly known by radio
observations of the molecular gas.  In between these hot
and cold phases, there is a warm (few hundred K) neutral medium and a
thermally ionized medium. ISO has made important contributions to our
understanding of the heating of the warm neutral gas and the
properties of the ionized gas in the GC.

Before ISO, the ionized gas has been mainly studied by radio continuum
and hydrogen radio recombination lines observations. They showed an
extended diffuse ionized gas component and a number of discrete \HII~
regions, most of them associated with the Sgr A, B,... complexes. Only
the most prominent \HII ~ regions (Sgr A) and ionized nebulae (the
Sickle) had been observed in infrared fine-structure lines. ISO
observations of fine-structure lines allowed studying the properties
of the ionized gas and the ionizing radiation over the whole GC
region. In the vicinity of Sgr A$^*$, where the ionizing source is the
central stellar cluster, \citeauthor{Lutz:96} (\citeyear{Lutz:96})
derived an effective temperature of the ionizing radiation of $\sim
35000$ K and \citeauthor{nemesio:01a} (\citeyear{nemesio:01a}) derived
a similar effective temperature in the Radio Arc region. They also
showed unambiguously that the Quintuplet and the Arches clusters
ionize the Sickle nebula and the thermal filaments that seem to
connect Sgr A to the Radio Arc. Indeed, the combined effects of both
clusters ionize a region of more that 40$\times$40 pc$^2$. Their
photo-ionization model simulations showed that, in order to explain
the long-range effects of the radiation, the ISM that surrounds the
clusters must be highly inhomogeneous. A similar scenario has been
invoked by \citeauthor{Goicoechea:04} (\citeyear{Goicoechea:04}) to
explain the extended ionized gas component in the Sgr B molecular
complex. Indeed, the whole GC region is permeated by relatively hot
($\sim 35000$ K) but diluted ionizing radiation field
(\citeauthor{nemesio:04a}, \citeyear{nemesio:04a}).

The line ratios (\NeIII/\NeII\ 15.5/12.8~\micron, \NIII/\NII\
57/122~\micron\ \ldots) measured in the GC are similar to those found
in low excitation starburst galaxies as M~83 or IC~342 but somewhat
lower than those measured in other starburst galaxies as NGC~3256 or
NGC~4945. However, the low ratios do not imply intrinsic differences
in the star formation activity as a lower upper mass cutoff of the
initial mass function. On the contrary, they are probably due to
starburst of short duration that produce hot massive stars but whose
age quickly softens the radiation (\citeauthor{Thornley:00},
\citeyear{Thornley:00}). The ISO observations of the GC are
consistent with a burst of star formation less than 7 Myr ago.

%%%%%%%%%%%%%%%%%%%%%%%%%%%%%%%%%%%%%%%%%%%%%%%%%%%%%%%%%%%%%%%%%%%%%%%%%
\begin{figure}[t!]
\includegraphics[width=\textwidth]{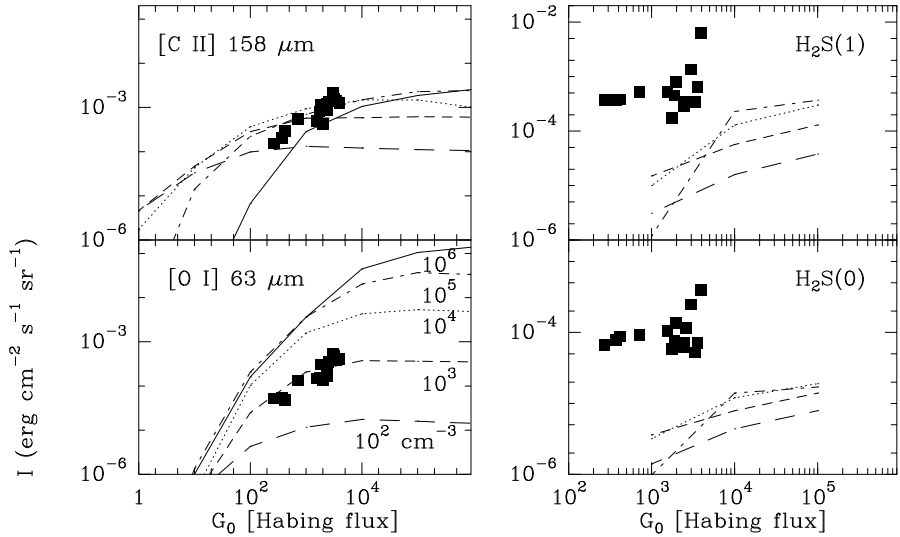}
\caption{Intensities of two fine-structure lines and two H$_2$
pure-rotational lines versus the far-ultraviolet incident flux
derived for the sources observed by Rodr{\'{\i}}guez-Fern{\' a}ndez et al.
(2004b). For comparison, the PDR model predictions
(Tielens \& Hollenbach, 1985) for different densities are also shown.}
\label{gal}
\end{figure}
%%%%%%%%%%%%%%%%%%%%%%%%%%%%%%%%%%%%%%%%%%%%%%%%%%%%%%%%%%%%%%%%%%%%%%%%%

On the other hand, in the GC there is a widespread neutral gas
component with temperatures of $\sim 200$ K and without associated
warm dust. Before ISO, this warm neutral gas had only been studied by
radio observations of symmetric top molecules like NH$_3$. How this
gas is heated has been a long-standing problem.  Radiative heating
mechanisms are usually ruled out due to the apparent lack of continuum
sources and the discrepancy of gas and dust temperatures.  ISO was
perfectly suited to study the heating of this gas. First,
\citeauthor{nemesio:01b} (\citeyear{nemesio:01b}) have measured for
the first time the total amount of warm gas in the GC clouds by
observing H$_2$ pure rotational lines. These lines trace gas with
temperature from 150 K to 500 K. The column density of gas with a
temperature of 150 K is a few 10$^{22}$ \cmmd\ and, on average, it is
around 30$\%$ of the total gas column density. Second, ISO has
observed the main coolants of the neutral gas (the \CII 158 and \OI 63
\mum\ lines) with temperatures of a few hundred K and the full
continuum spectrum of the dust emission around the maximum.  The dust
emission can be characterized by two temperature components, one with
a temperature of $\sim 15$ K and a warmer one with a temperature that
varies from source to source from $\sim 25$ to $\sim 45$ K
(\citeauthor{lis:01}, \citeyear{lis:01}; \citeauthor{nemesio:04b},
\citeyear{nemesio:04b}; \citeauthor{Goicoechea:04}
\citeyear{Goicoechea:04}). A detailed comparison of the lines and
continuum emission with shocks and PDR models show that the fine
structure lines and the H$_2$ emission from excited levels arise in a
PDR with a far-ultraviolet incident radiation field 10$^3$ times
higher than the average local interstellar radiation field and a
density of 10$^3$ \cmmt (\citeauthor{nemesio:04b}
\citeyear{nemesio:04b}).  The warm dust emission and the
fine-structure lines probably arise in PDRs located in the interface
between the ionized gas discussed above and the cold molecular gas.
\citeauthor{nemesio:04b} (\citeyear{nemesio:04b}) also showed that an
important fraction ($\sim 50 \%$) of the \CII 158 and \SiII 35 \mum\
can arise from the diffuse ionized gas instead of the PDR
itself. Regarding the warm H$_2$, only 10-20 $\%$ of the total column
density of gas with temperature of 150 K can be accounted for in the
widespread PDRs scenario. The most likely heating mechanism for the
bulk of the warm neutral gas is the dissipation of
magneto-hydrodynamic turbulence or low velocity shocks.

Spectroscopic imaging/mapping will be useful to investigate the
interplay between the different ISM phases in the GC. This could be
done with the IRS instrument onboard the Spitzer Space Telescope. In
addition, SOFIA and Herschel will be able to map the ionized and the
warm neutral gas with high spectral resolution, which is needed
due to the crowded velocity fields in the GC region.

\section{Supernova Remnants (SNRs)}
\label{sn}

%%%%%%%%%%%%%%%%%%%%%%%%%%%%%%%%%%%%%%%%%%%%%%%%%%%%%%%%%%%%%%%%%%%%%%%%%
\begin{figure}[t!]
\includegraphics[clip,width=\textwidth]{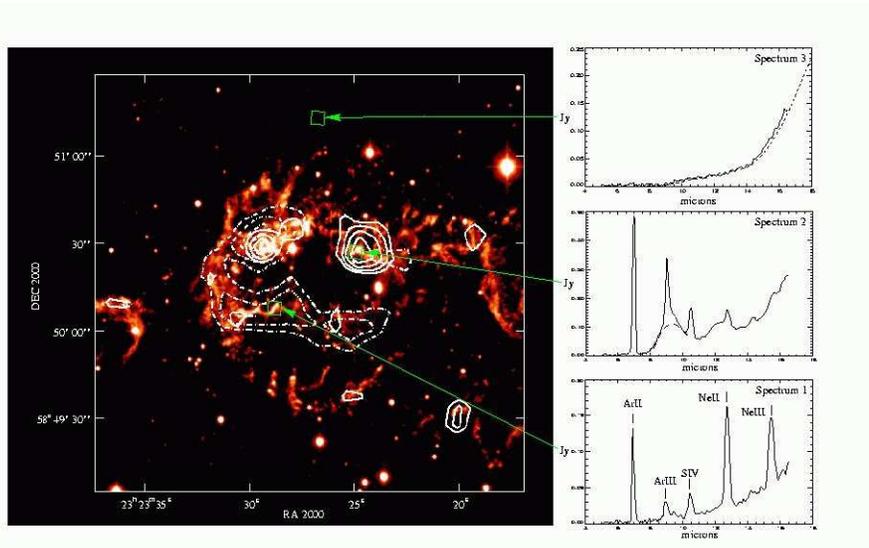}
\caption{Contour maps of the \NeII\ 12.8~\micron\ line emission
(dotted contours) and of the 9.5 \mum\, silicate dust emission (full
contours) plotted onto an optical image of the Cassiopeia-A
supernova remnant. Examples of the variety seen in the MIR spectra are
shown on the right [Taken from Douvion et al. 1999].}
\label{snr}
\end{figure}
%%%%%%%%%%%%%%%%%%%%%%%%%%%%%%%%%%%%%%%%%%%%%%%%%%%%%%%%%%%%%%%%%%%%%%%%%

% Cas A

As the prototype of O-rich young SNR, Cassiopeia A is well-studied by
ISO \cite{Lagage, Tuffs, Arendt, Douvion:99, Douvion}. Several
fine-structure lines are observed at high velocity with [OIV] being
the most prominent (e.g. Fig.\ref{snr}). However, no iron emission is
observed \cite{Arendt}. These fine-structure lines are spatially
correlated with the Fast Moving Knots (FMKs) known to be made of
nuclear burning products from the progenitor star. While fainter
diffuse dust emission is present, the brightest continuum emission is
spatially correlated with the fine-structure lines indicating that
dust has freshly condensed in the FMKs. The high dust temperature
derived is consistent with emission from collisionally heated dust in
the post-shock region of FMKs. A new 22 \mum\, feature is found and
identified with Mg protosilicates \cite{Arendt}. However, Mg
protosilicates give rise to a 10 \mum\, feature quite different from
the observed one \cite{Douvion}. These authors fitted a 6--30 \mum\,
spectrum of Cas A with SiO$_2$, MgSiO$_3$, Al$_2$O$_3$. In addition to
dust continuum emission, silicate emission is present at several
positions within Cas A (Fig.\ref{snr}). The presence of silicate and
the presence of neon is anti correlated in many knots while argon and
sulfur are present in most of the neon knots (Fig.\ref{snr}),
suggesting a different degree of mixing occurred
\cite{Douvion:99}. These results also indicate an incomplete
condensation of silicate elements. If born out for other core collapse
SNe, these SN can no longer be considered the dominant source of
silicates \cite{Douvion:99}. Similar conclusion was reached by
\citeauthor{Tuffs} (\citeyear{Tuffs}) based upon the derived dust mass
being less than a few hundredths of a solar mass.

In contrast to Cas A, the dust emission in the Kepler and Tycho SNR is
spatially correlated with H$\alpha$ emission and therefore attributed
to shocked circumstellar (CS) material and CS or interstellar (IS) material
respectively and not to SN condensates. Instead of dust signatures,
the Crab SNR is dominated by synchrotron radiation with a spectral
index of -0.3 to -0.65.  Fine-structure lines of \NeII\, and \ArII\,
are observed for the Kepler SNR and of \NeII, \NeIII, \NeV, \ArII,
\ArIII, \SIV, \NiII\, for the Crab SNR. The \NeIII\, emission in the
Crab SNR shows the same filamentary structure as seen in optical
lines, known to be composed predominately of SN
ejecta \cite{Douvion:tycho}.

The IR spectrum of RCW103, a young and fast SNR, is dominated by
prominent lines from low-excitation species allowing estimates of
electron density, temperature and abundances. The ionic lines are
consistent with post-shock emission. The absence of significant
emission from the pre-shock region suggests that shock models may
overestimate the importance of the precursor. All - but one -
abundances are solar and hence the emitting gas is ISM material in
which the dust grains have been destroyed by the shock front
\cite{Oliva:rcw}.

The interaction of SNR with molecular clouds has been extensively
studied with ISO.  The IR spectra of the SNRs 3C~391, W44 and W28
exhibit fine-structure line emission, dust emission along the line of
sight and shocked dust emission shortwards of 100 \mum\, (Reach \&
Rho, 1996, 2000; \citeauthor{Reach:02}, \citeyear{Reach:02}). In
addition, shock-excited FIR emission of H$_2$O, CO and OH emission is
detected in 3C~391 \cite{Reach:98}. These IR observations provide
evidence for multiple pre- and post-shock conditions which are found
to be spatially separated. Indeed, ISOCAM-CVF observations clearly
resolves the ionic emission from the molecular emission tracing
moderate-density to high density regions in 3C~391. Abundance analysis
reveal partial dust destruction consistent with theoretical models of
grain destruction. Similar, spectral variations are observed spatially
within IC443 \cite{Cesarsky, Oliva, Rho}. The north-easthern
filamentary structure is dominated by prominent \NeII, \FeII, \SIII,
\OIV \, and \OI\, line emission while the southern clumpy region is
dominated by H$_2$ emission indicating different types of shock
occuring in different conditions. Both regions can alone account for
virtually all the observed IRAS flux in the 12 and 25 \mum\, bands --
likely due to the limited ISO field-of-view and limited IRAS spatial
resolution. Despite this, the unusually blue IRAS colors of IC443
reflect line (molecular and/or atomic) contamination instead of a
large population of small grains.

The late emission of SN1987A probes directly the elemental abundances
deep in the stellar ejecta and as such constrains models of SN
explosion and the explosive nucleosynthesis. Particularly important
for this are the $^{56}$Ni, $^{57}$Ni and $^{44}$Ti masses. The former
two are known, the latter is more uncertain and can be derived from
the \FeI24.05 \mum\, and \FeII25.99 \mum\, fine-structure lines.
These lines are however not detected in the ISO spectra of SN1987A
indicating -- based upon time dependent models -- an upper limit of
1.1 10$^{-4}$ M$_\odot$ (Lundqvist et al. 1999; 2001).  The central
region around the SN position is resolved in the MIR and consistent
with collisionally heated grains in the shocked CS gas around SN1987A
\cite{Fischera:02}. The MIR spectrum can be modeled by a dust mixture
of silicate-iron or silicate-graphite or pure graphite grains and
indicates a low dust-to-gas ratio \cite{Fischera:02}.

Hence, observations of ionic lines allow to diagnose the physical
conditions of the medium in which the SNR blast wave is propagating,
the shock kinematics, the elemental abundances and hence also dust
destruction. ISO provided evidence that dust can form in supernovae
and dust can be (partially) destroyed in supernovae.

\acknowledgements

EP acknowledges the support of the National Research Council. NJR-F
acknowledges support by a Marie Curie Fellowship of the European
Community under contract number HPMF-CT-2002-01677.

\end{article}
\end{document}